# Machine Learning based Optimal Feedback Control for Microgrid Stabilization

Tianwei Xia[1], *Student Member, IEEE*, Kai Sun[1], *Senior Member, IEEE*, Wei Kang[2], *Fellow, IEEE*
[1]Department of Electrical Engineering and Computer Science, University of Tennessee, Knoxville, TN, USA
[2]Naval Postgraduate School, Monterey, CA, USA
txia4@vols.utk.edu, kaisun@utk.edu, wkang@nps.edu

*Abstract*— Microgrids have more operational flexibilities as well as uncertainties than conventional power grids, especially when renewable energy resources are utilized. An energy storage based feedback controller can compensate undesired dynamics of a microgrid to improve its stability. However, the optimal feedback control of a microgrid subject to a large disturbance needs to solve a Hamilton-Jacobi-Bellman problem. This paper proposes a machine learning-based optimal feedback control scheme. Its training dataset is generated from a linear–quadratic regulator and a brute-force method respectively addressing small and large disturbances. Then, a three-layer neural network is constructed from the data for the purpose of optimal feedback control. A case study is carried out for a microgrid model based on a modified Kundur's two-area system to test the real-time performance of the proposed control scheme.

## I. Introduction

With the fast development of power electronic converters in recent years, more and more distributed energy resources such as renewable generators are utilized to provide electricity supplies to customers connected to a conventional distribution network or direct users of a microgrid [1]. At present, microgrids attract a lot of interests from both academia and industry because they can be easily deployed to serve users of a local area and are more flexible in adopting emerging technologies in smart grid and control systems. A microgrid can accommodate a large variety of energy resources, especially power electronics-interfaced renewable generators [2]. Thus, the control design for a microgrid is more flexible than the control of a conventional power grid and poses new challenges in operations due to the dynamics of the renewable generators. For instance, power oscillations locally in a microgrid may have faster frequencies and behave with more uncertain properties [3]. Any energy resource connected through a power converter may become a source of oscillation to impact the microgrid and even propagate toward the main power grid connected.

An energy storage system (ESS) can play a critical role in complementing the uncertainties with renewable generators [4]. It has been used to improve load-following by operating in a flexible switchable charging or discharging mode. Many researchers have investigated the control of the ESS to improve the power system stability [5, 6]. One obvious feature is that the ESS can significantly increase the damping of oscillation if well-designed, especially in a weak power grid [7, 8].

This paper focuses on the optimal feedback control problem about the ESS in a microgrid. Such a nonlinear optimization problem can be tackled by solving a Hamilton-Jacobi-Bellman (HJB) equation, a partial differential equation (PDE). For this problem, it is hard to find the globally optimal solution due to the non-convexity of the problem and the curse-of-dimensionality [9]. In the past two decades, many researchers have studied approximate solutions to optimal feedback control problems in other engineering fields [10, 11]. Unfortunately, the time performance of most methods is not satisfying for large systems, which is critical to ESS control to ensure real-time operations of microgrids.

In the literature, the Machine learning method has been extensively applied to solve many difficult engineering problems nowadays [12]. If trained well with sufficient data, a machine learning tool can meet the real-time performance requirement of control. This paper proposes a machine learning-based optimal feedback control approach, which is used to design an ESS-based optimal feedback controller for a microgrid subject to small and large disturbances. The rest of the paper is organized as follows: Section II introduces the proposed machine learning-based optimal feedback control approach. The microgrid model and two numerical approaches are discussed in detail. Besides, some measures are proposed to improve the performance of training. The case studies on a microgrid model connected to a main power grid model are presented in section III. At last, conclusions are drawn in section IV.

## II. Approach

This section first introduces the objective function for the optimal control in *A* and system models in *B*. Two mature methods, LQR and Brute-force search, are reviewed in *C* and *D*, respectively. The interpretation of the proposed machine learning is shown in *E* in detail.

This work was supported in part by the ERC Program of the NSF and U.S. DOE under grant EEC-1041877 and in part by the NSF grant ECCS-1553863.

## A. Hamilton-Jacobi-Bellman Problem

For nonlinear systems, optimal feedback control is designed based on the solution of the HJB equation [11]. The fixed final time-optimal control problem can be formulated as follows:

$$\min_{u \in U} \quad F(\mathbf{x}(t_f)) + \int_0^{t_f} L(t, \mathbf{x}, \mathbf{u}) dt$$
$$s.t. \quad \dot{\mathbf{x}}(t) = f(t, \mathbf{x}, \mathbf{u}) \quad (1)$$
$$\mathbf{x}(0) = \mathbf{x}_0$$

where the $\mathbf{x}$ is the state variable, and $\mathbf{u}$ is the control variable. $F$ and $L$ are the terminal cost and running cost, respectively. $f$ is a Lipschitz continuous vector field, and it is also the description and constraints of the system. The given initial state is $\mathbf{x}_0$ (avoid initiating a sentence with a math notation). The optimal solution is denoted by

$$\mathbf{u} = \mathbf{u}^*(t; \mathbf{x}_0) \quad (2)$$

Thus, the value function can be written by

$$V(t, x) = \inf \left\{ F(\mathbf{y}(t_f)) + \int_0^{t_f} L(t, \mathbf{y}, \mathbf{u}) dt \right\} \quad (3)$$

## B. Power System Model

A practical microgrid is often equipped with one or more synchronous generators, such as a diesel generator for ease of control, or alternatively renewable generators that are also able to emulate synchronous generators with synthetic inertia. Thus, the paper uses a small synchronous generator as described by the swing equations in (4) to model a renewable generator in the microgrid and then formulate the optimal feedback control problem without loss of generality.

$$\frac{d\delta_i}{dt} = \Delta\omega_i$$
$$\frac{2H_i}{\omega_0} \frac{d\Delta\omega_i}{dt} = P_{mi} - P_{ei} - K_{Di} \frac{\Delta\omega_i}{\omega_0} \quad (4)$$

where $\delta_i$ is the rotor angle, and $\omega_i$ is the rotor angle speed of generator $i$, respectively. $H_i$ and $K_{Di}$ is the inertia and damping coefficient of the machine of the generator $i$. $P_{mi}$ and $P_{ei}$ are mechanical power and electromagnetic power of generator $i$, respectively. $\omega_0$ is the rated angular frequency.

The proposed controller adjusts the level of injected power from the ESS to stabilize the microgrid under disturbances. Thus, the HJB problem for a synchronous generator connected to an ESS is

$$\min_{u \in U} \quad v = F(\mathbf{x}(t_f)) + \int_0^{t_f} L(t, \mathbf{x}, \mathbf{u}) dt$$
$$s.t. \quad \dot{\mathbf{x}}(t) = f(t, \mathbf{x}, \mathbf{u}), \quad \mathbf{x}(0) = \mathbf{x}_0 \quad (5)$$
$$L(\boldsymbol{\delta}, \Delta\boldsymbol{\omega}, \mathbf{u}) = w_1 \|\boldsymbol{\delta} - \boldsymbol{\delta}_s\|^2 + w_2 \|\Delta\boldsymbol{\omega}\|^2 + w_3 \|\mathbf{u}\|^2$$

where

$$\mathbf{u} = \mathbf{P}_{control} \quad (6)$$

$\mathbf{u}$ is the power injection and control variable of the system. The variable $\boldsymbol{\delta}_s$ is the final state vector of the rotor angle. Since the system needs to be stable with control, the final state of rotor angle speed $\boldsymbol{\omega}_s$ should be equal to the synchronous frequency with a zero deviation. $w_1$ to $w_3$ are the weights of three state variables, whose values depend on the unit and the range of each state variable.

## C. Approximation LQR Solution

Equation (5) is a typical nonlinear optimal feedback control problem. It is almost impossible to find a general, analytical solution of this HJB equation [10, 11]. However, under a small disturbance, the system oscillates around the stable equilibrium so that the system can be approximated by a linear system. In such a way, the problem is simplified to a linear–quadratic regulator (LQR) problem as described by

$$\dot{\mathbf{x}} = \mathbf{A}\mathbf{x} + \mathbf{B}\mathbf{u}$$
$$J(\mathbf{u}) = \int_0^\infty (\mathbf{x}^T \mathbf{Q}\mathbf{x} + \mathbf{u}^T \mathbf{R}\mathbf{u} + 2\mathbf{x}^T \mathbf{N}\mathbf{u}) dt \quad (7)$$
$$\mathbf{u} = -\mathbf{K}\mathbf{x}$$

where $\mathbf{A}$ and $\mathbf{B}$ are the state matrix and input matrix, $J$ is the quadratic cost function, $\mathbf{Q}$, $\mathbf{R}$, and $\mathbf{N}$ depend on the practical system, and $\mathbf{K}$ is the coefficient of the controller.

The LQR problem has been studied for many years, and there are several mature tools to solve it efficiently [13]. Unfortunately, this solution is only valid in a small neighborhood of the equilibrium in the state space, where the system can be regarded as a near-linear system.

## D. Brute-force Search

Given a large disturbance, the system's nonlinearity is significant and should not be neglected anymore. Therefore, the performance of the LQR solution cannot be guaranteed, so a nonlinear control problem has to be solved. The brute-force search (BFS) method is a simple, yet time-consuming offline method to find optimal control of a general nonlinear system by, e.g., the following three steps: first, the control period of interest is divided into short intervals; second, for each time interval, the control variable, i.e., the value of the power injection from the ESS, is set to be a constant value from a given range; third, the system performance with each combinational control over all intervals is checked in order to identify the optimal curve of control over the entire time period. The optimality of the control determined by the BFS method depends on the length of intervals. Shorter intervals will generate controls closer to the true optimal control but will incur the curse-of-dimensionality.

## E. Proposed Machine Learning-Based Approach

As discussed in *C* and *D*, the LQR method is fast but inaccurate under a large disturbance; the BFS method can provide a relatively accurate result for a large disturbance, but the time performance is not acceptable if the time period of control is partitioned into many time intervals. Both methods are applied in the proposed machine learning (ML) based approach to generate training data respectively under small and large disturbances: for offline simulated disturbances, the BFS is used to find optimal feedback control against large disturbances while the LQR control is applied to small disturbances. Thus, the two methods can be integrated into a hybrid approach to generating the training dataset.

Fig. 1 and Fig. 2 provide an example for the illustration of the system responses with control schemes determined respectively by LQR and BFS methods. The response of the system without control is shown in Fig. 1. The blue dotted curve in Fig. 2 is the result of the optimally controlled power injection of ESS of the LQR approach, while the BFS result is in the red dashed curve. Because the time length for each time interval is large, the optimal control from the BFS is a decaying square wave rather than a relatively smooth waveform. The nonlinearity of the system is not obvious due to the small disturbance, so the two trajectories are close to each other.

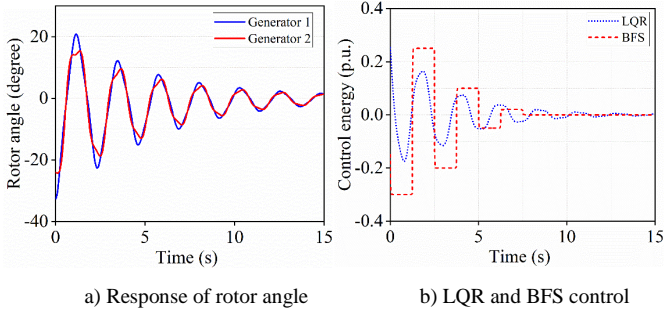

a) Response of rotor angle  b) LQR and BFS control

Fig. 1. A typical response and the optimal control

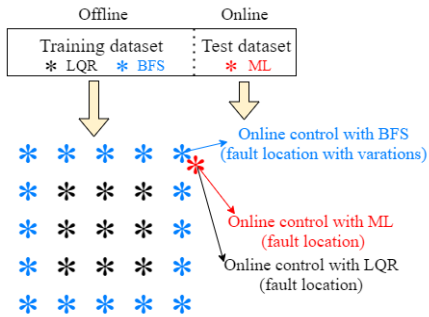

Fig. 2. Offline and online training datasets

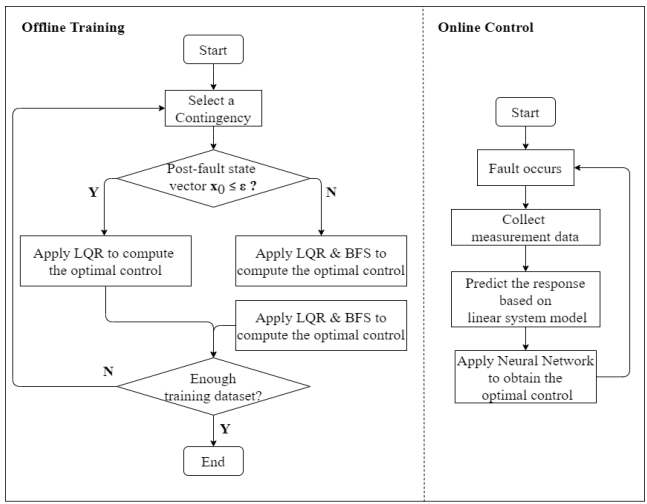

Fig. 3. Flowchart of the machine learning approach

It is known that the BFS method provides a reliable result if the length of the time interval is small enough. Its advantage lies in the validity of a general nonlinear system. However, the computation burden will become extremely large in order to get a relatively accurate result.

As long as the control scheme of the ESS is determined, the response of the system will be obtained by computer simulation on the control scheme. Then the cost of each control scheme is calculated by equation (3). Different control schemes will result in different cost values, and the control with the minimal cost will be the optimal control. In this way, the training dataset can be generated by the hybrid of BFS and LQR. The practical case study shows that for the response from a small disturbance, the LQR is selected while the large one, the BFS will be the optimal method. The flowchart of the offline training is shown in the left part in Fig. 3.

Note that the training data set may not cover all possible contingencies, so the variations need to be made on the data to the robustness of control. It is feasible for the LQR and ML methods to apply real-time control with the fault location information(red star in Fig. 2). However, for the BFS, due to the relative worse time performance, it is almost impossible to compute the optimal control online; thus, the optimal control with the closest location (blue start in Fig. 2), which is computed offline, will be chosen as the optimal control for BFS method. By this way, the BFS, LQR and ML online optimal control can be compared fairly. The details for the test set will be discussed in the case study in section III.

It should be emphasized that all the training and testing tasks are conducted offline. Ideally, the inputs to the neural network are rotor angles and speeds of generators over 15-second time windows, as shown in Fig. 1a). However, in the real-time operating environment, a microgrid cannot wait, e.g., 15 seconds for the input data and then make a decision of control. In fact, control should be functioning all the time based on online data.

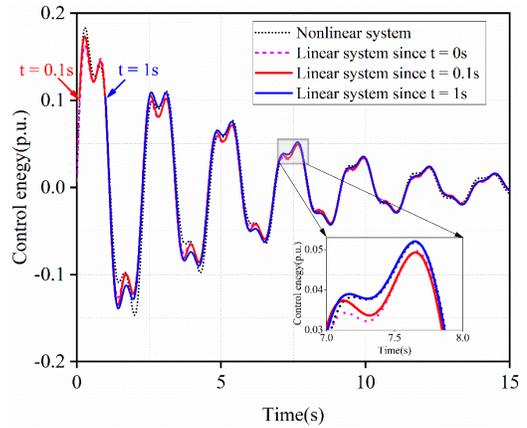

Fig. 4. Illustration for the real-time control

Fortunately, the response of the microgrid in a whole 15 seconds time period is approximately predictable from its response in the first 0.1-1 second time interval. Especially when the system is subject to a small disturbance, its response over the entire 15 seconds can be analytically solved. Thus, it is feasible to predict the system's response for the entire time window using a measured system response during a sliding 0.1-1 second short time window under both small and large disturbances. The predicted system response is then used as input data to determine the optimal control from the neural network. Fig. 3 shows an example of the system response and the predicted system response. The black dotted curve denotes the response of the nonlinear system, which is also the true response. The red solid and blue solid represent the responses of the linear system with 0.1 and 1-second measurement data, respectively. The value of control given by the neural network for the current time instant is executed on the ESS. By using this algorithm, real-time control is ensured. The right part of Fig. 4 shows the flowchart of the online control.

### III. CASE STUDIES ON POWER SYSTEM

#### A. System Topology

A modified Kundur's two-area system is used for the case study in this paper, which is shown in Fig. 5. One area is used to model the microgrid, and the other one represents the main grid. There are two generators in the microgrid. Generator 1 is the diesel generator, and generator 2 is a solar farm operated as an emulated synchronous generator with synthetic inertia. A battery-based ESS is added to the microgrid to provide optimal feedback control.

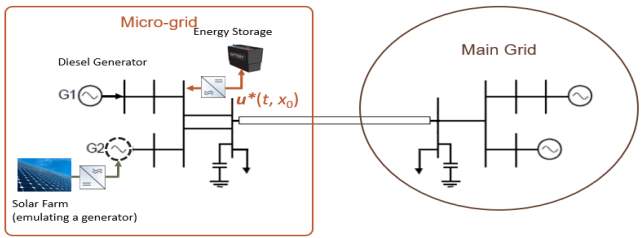

Fig. 5. The structure of the microgrid system.

#### B. Optimal Control with LQR and BFS

Both the LQR and BFS methods are applied to find optimal control under small and large disturbances, respectively. TABLE I compares the costs of different scenarios: without control or with control optimized by the LQR or BFS method. There is no doubt that the cost with control is smaller than the cost without control, which means that the ESS can stabilize the response of the system. Under small disturbance, the system can be viewed as a linear system, and the control from LQR is about the same as that from the BFS or even slightly better. The length of the time window for BFS is 1.25 seconds, and it can also solve the optimal control for the linear system as long as the time window is small enough.

TABLE I. THE RESULT FROM LQR AND BFS METHODS

| Cases | | LQR | BFS |
|---|---|---|---|
| Small disturbance | Without control | 0.0041 | 0.0042 |
| | With control | 0.0029 | 0.0034(1.25s) |
| Large disturbance | Without control | - | 0.3045 |
| | With control | - | See TABLE II |

Under a large disturbance, the system does not behave linearly. The cost of control defined in (5) is calculated for the BFS method using different time intervals. The results are compared in TABLE II, which are all lower than the cost of 0.2775 with the result of the LQR method because of the nonlinear response of the system under the large disturbance.

TABLE II. RESULT FROM BFS WITH DIFFERENT INTERVALS

| Interval | 1.25s | 1.00s | 0.50s | 0.35s |
|---|---|---|---|---|
| Cost | 0.2595 | 0.2467 | 0.2392 | 0.2384 |

Based on this case study, the control from the LQR method is not optimal under a large disturbance. The control from the BFS method is better but cannot be applied in real-time. Machine learning is then used to accelerate the speed of finding the optimal control in real-time for a large disturbance based on offline training.

#### C. Training and Test Dataset

The training set includes 6722 cases generated by both methods. First, as illustrated in Fig. 6, two typical fault locations (with red and blue responses) with gradually increased sizes of faults are considered to create a number of disturbances for the BFS method to find optimal controls. For the post-fault trajectory of each disturbance, any point on it can be regarded as a new initial state. Thus, total 3,000 cases are generated on large disturbances. The other 3,731 cases come from the LQR method for evenly distributed initial states in a small range deviated from two rotor angles of the equilibrium, i.e., -15 to 15 degrees at 0.5 degree intervals, as illustrated by the square region in Fig. 6.

As for the test data, 10% and 20% variations (as illustrated by the round regions in Fig. 6) are made on selected large disturbances to address uncertainties of the disturbances in practice. In order to compare different approaches fairly, the control from BFS should come from the off-line training set, while the control from LQR and machine learning can be computed based on the online measurement.

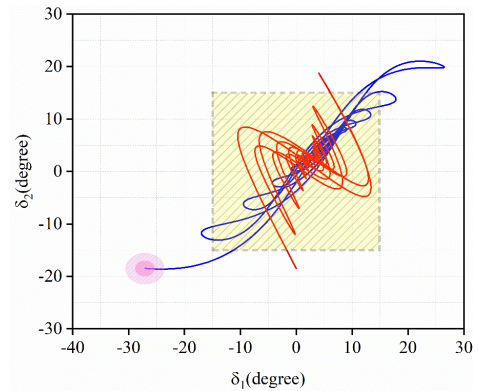

Fig. 6. The distribution of the training and test dataset.

#### D. Performance of Machine Learning

The neural network used in this paper has three hidden layers of 64 ReLU units. The input is a time series of four state values at a 100Hz sampling rate, and the output is the optimal control from the LQR or BFS method. Fig. 7 shows results from the LQR, BFS (1-second interval) and neural network responding to the same disturbance.

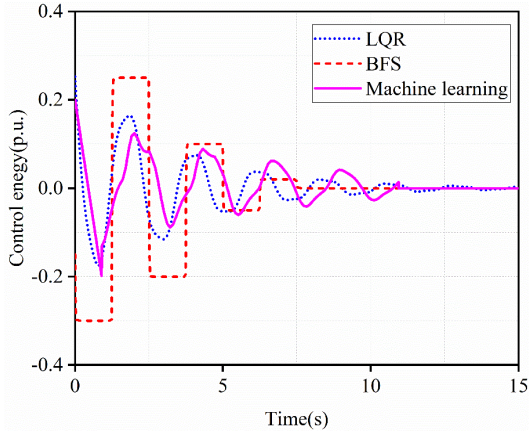

Fig. 7. The optimal control from different approaches.

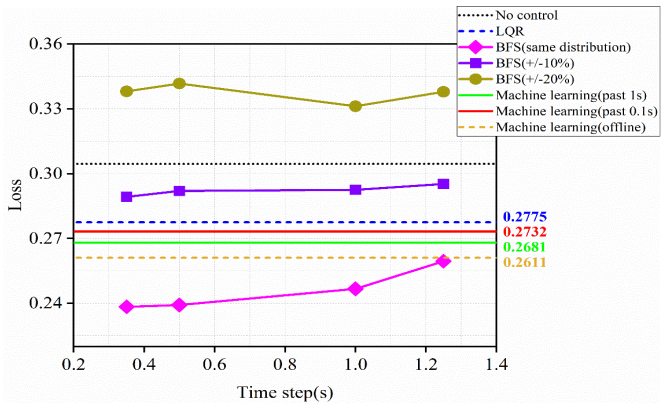

Fig. 8. The cost of different approaches.

The control costs of three different methods are compared in Fig. 8. The curves on the BFS method give how the cost of the BFS method can change with the length of the time interval for disturbances with and without variations. The offline machine learning approach has a satisfying performance because the input data is 15 seconds system response. The online machine learning approach is a little worse since the input trajectory is predicted based on the first 0.1 or 1 second. However, under large disturbance, the online machine learning approach is still better than the LQR and BFS method with a variation.

LQR has poor performance on large disturbances, but it can be used to generate training data for small disturbances or when the system becomes close to the equilibrium after a large disturbance. For each disturbance, the BFS method takes 6 minutes to find the optimal control varying at 1-second intervals, which requires about 60,000 numerical simulations on the microgrid. It is not robust against small variations of the disturbance.

Machine learning takes 5 minutes for offline training using the results from both the BFS and LQR methods, and takes less than 1 second to give real-time control. Although the control from machine learning is not as good as that from the BFS method when no variation is considered on the disturbances, machine learning is overall more robust against variations of the disturbance. The training data set is critical for the performance of the machine learning-based method. In the practical system, with the operation of a microgrid system, more and more data are collected and computed from online measurement and offline simulation. The result from the machine learning method will become more accurate and robust, so the upgradability is one of the most critical advantages compared with other traditional methods.

IV. CONCLUSION

This paper has proposed a machine learning-based optimal feedback control approach for microgrid stabilization under small and large disturbances. Two different methods, the LQR and BFS methods, are used to generate the training dataset. From a case study on a microgrid model, the proposed machine learning-based approach has the potential to provide the optimal feedback control in real-time for a microgrid by means of energy storage devices.